# High-resolution synchrotron XRD study of Zr-rich compositions of $Pb(Zr_xTi_{1-x})O_3$ (0.525≤x≤0.60): evidence for the absence of the rhombohedral phase

Akhilesh Kumar Singh and Dhananjai Pandey[a)]
School of Materials Science & Technology, Institute of Technology,
Banaras Hindu University, Varanasi-221 005, INDIA

Songhak Yoon and Sunggi Baik
Department of Materials Science and Engineering,
Pohang University of Science and Technology, Pohang 790-784, Korea

Namsoo Shin
Pohang Accelerator Laboratory, Pohang University of Science and Technology,
Pohang 790-784, Korea

## Abstract

Results of Rietveld analysis of the synchrotron XRD data on $Pb(Zr_xTi_{1-x})O_3$ (PZT) for 0.525≤x≤0.60 are presented to show the absence of rhombohedral phase on the Zr-rich side of the morphotropic phase boundary (MPB). Our results reveal that the structure of PZT is monoclinic in the Cm space group for 0.525≤x≤0.60. The nature of the monoclinic distortion changes from pseudo-tetragonal for 0.525≤x≤0.54 to pseudo-rhombohedral for x>0.54.

[a)] Author to whom correspondence should be addressed;
electronic mail: dpandey@bhu.ac.in

The phase diagram of the solid solution of $PbTiO_3$ with several other oxide perovskites contains a nearly vertical morphotropic boundary (MPB) for which the piezoelectric and dielectric responses show extremum response [1]. The origin of this extremum response is under intense debate in recent years. The first MPB ceramic discovered and used extensively in several actuator and sensor devices is $Pb(Zr_xTi_{1-x})O_3$ (PZT) [1]. The MPB in PZT has all along been believed to separate the tetragonal (T) and rhombohedral (R) phase regions separated by a two phase region whose intrinsic width in chemically homogeneous and stoichiometric samples is as small as 0.010 [2] although much higher widths have been reported for less homogeneous samples prepared by the solid state route (e.g. $\Delta x \approx 0.15$ in Ref.3 and 0.05 in Ref.4, 5). It was predicted by Mishra et al [2] that the phase coexistence in the narrow ($\Delta x \approx 0.010$) MPB region could be due to a phase transition from a low temperature phase stable below room temperature to the T phase stable above room temperature. This was indeed confirmed recently for tetragonal compositions close to the MPB (x=0.520), where a low temperature monoclinic phase with Cm space group is found to be the stable phase [6]. This is an $M_A$ type monoclinic phase (in the notation of Ref.7) which transforms further to a superlattice phase first reported by Ragini et al [8] and subsequently confirmed by others [9]. The space group of this superlattice phase was shown to be Cc by Hatch et al [10] which was confirmed later on by others too [11]. The stability of the Cm and Cc phases has been confirmed in *ab initio* first principles calculations [12]. The Cc phase has also been reported in the high pressure studies [13]. These results have been reviewed recently [14].

The polarization vector of the $M_A$ phase with Cm space group can lie anywhere between [001] polarization direction of the T phase and the [111] direction of the R phase. As a result, the $M_A$ phase in the MPB region has been postulated as a bridge between the tetragonal and rhombohedral phases on the two sides of the MPB [4, 6, 15]. We present here the results of Rietveld analysis of the powder synchrotron XRD data which reveals that the structure of PZT for x=0.525 is monoclinic of $M_A$ type, and not a mixture of (a) T and R [1-2], (b) coexistence of $M_A$ and T [4,16] and (c) coexistence of R and $M_A$ [17], proposed by the pervious workers. More significantly, our results suggest the absence of any rhombohedral phase on the Zr-rich side ($x \geq 0.525$) of the MPB raising doubts about the applicability of the widely believed notion of $M_A$ structure as the bridging phase between T and R phases. Our analysis reveals that the nature of the monoclinic distortion of the Cm phase changes from pseudo-tetragonal for $0.525 \leq x \leq 0.540$ to pseudo-rhombohedral for higher Zr-content.

In order to determine the phase stabilities of the PZT near MPB, which is very sensitive to compositional variations, it is essential to have samples of exceptionally high quality in terms of chemical homogeneity and stoichiometry. Our samples were prepared by a semi-wet route [18] which is known to give the narrowest width ($\Delta x \approx 0.010$) of the MPB region. For powder diffraction measurements, sintered pellets were crushed to fine powders and annealed overnight at 773 K to get rid of strains introduced during crushing. Synchrotron powder XRD experiments were carried out at 8C2 HRPD beamline at Pohang Light Source (PLS), Pohang Accelerator Laboratory, Pohang, Korea. The incident x-rays were monochromatized to the wavelength of 1.543 Å by a double bounce Si (111) monochromator. The diffraction data was collected in the 10 to 130 ($2\theta$) degree range at a step of 0.01degree. Rietveld analysis of the data was carried out using Fullprof-Suite [19].

Fig.1 depicts the 200, 220 and 222 reflections of PZT with different Zr content. The doublet character of 200 and 220 and singlet nature of 222 confirms tetragonal structure for x=0.515. The structure is 'rhombohedral' for x>0.525, as 200 is now a singlet while 222 is a doublet. For x=0.525, the 200 and 222 peaks are split which rules out the T or R structures for this composition. We carried out full pattern Rietveld refinements for x=0.525 using all possible models proposed in the literature: (a) coexistence of T and R [1-2], (b) coexistence of R and Cm [17] (c) coexistence of Cm and T [4,16] and (d) pure Cm phases. The results for a few selected profiles are shown in Fig.2. It is evident from Fig.2 that Cm space group accounts very well for the observed profiles for this composition. The refinement for the coexistence of T and R phases leads to very poor fit with much higher $\chi^2$. Consideration of a coexisting T phase with the monoclinic phase although decreases the $\chi^2$ from 2.08 to 1.93, but it is statistically not significant enough considering the increase in the number of refinable structural parameters from 15 to 24. The consideration of coexisting R phase with Cm, on the other hand, deteriorates the fits as evidenced by the increase in the $\chi^2$. Our results thus, clearly establish that the structure of PZT for x=0.525 is monoclinic of $M_A$ type in the Cm space group. On increasing the Zr-content (x>0.525), the splitting of the 200 peak in Fig.1 disappears but there is an anomalous broadening of this peak. Fig.3 shows the composition dependence of the ratio of the full widths at half maximum (FWHM) of 200 and $1\ 1\ \bar{1}$ reflections (using rhombohedral indices). This ratio is nearly equal to 1 for the tetragonal compositions, as expected. For the rhombohedral compositions also this ratio should have been nearly equal to one but it is much larger for x≥0.535. Such anomalous broadening is usually accounted for in the Rietveld refinements by considering anisotropic peak broadening functions [20]. However, the consideration of the anisotropic peak broadening functions led to mismatch in the observed and calculated profiles for other reflections like 111 [see Fig.4(a), (e)]. If we constrain the refinement to account for the observed 111 profiles, the mismatch appears for the h00 (e.g. 200) and hh0 (e.g. 220) [see Fig.4(b)]. Attempts to refine the structure using off-centre $<110>_{cubic}$ Pb displacement in the R3m space group, as proposed by Corker et al [21] for PZT compositions much richer in Zr-content, also did not improve the fits (see Fig.4(c)). All these comprehensively rule out the R3m space group for x=0.535. Use of Cm space group, on the otherhand, leads to much lower $\chi^2$ and excellent fits for both the 200 and 111 type reflections. These results thus clearly favour the Cm space group for x=0.535. Similar refinements for higher Zr-compositions revealed the inadequacy of the rhombohedral structure and the correctness of the Cm space group. This is illustrated in Fig.4 (e) and (f) for x=0.60. Consideration of a coexisting R3m phase with Cm for x=0.60 change $\chi^2$ from 1.67 to 1.65 but it is statistically insignificant considering the increase in the number of refinable structural parameters from 15 for pure Cm to 24 for Cm+ R3m. The fact that the anomalous broadening of the h00 and hh0 type reflections persists even beyond x=0.60 (see Fig.3) suggests the absence of rhombohedral distortion for x>0.60 also.

Fig.5(a) depicts the variation of the equivalent elementary perovskite cell parameters with composition at 300K for the Zr-rich compositions using Cm space group. The cell parameters of the $M_A$ phase for x=0.520, for which the 'T' and '$M_A$' phases coexist [22], have also been included in Fig.5. It is evident from Fig.5(a) that the $a_m$ and $b_m$ lattice parameters initially increase with composition and then show saturation for x≥0.545. The monoclinic angle β also increases with increasing 'x' upto x< 0.545 but starts decreasing for x≥0.545. The value of $c_m$ is highest for x=0.520 and it decreases rather fast with x before showing saturation for x≥0.535. Fig.5(b) shows the variation of unit cell volume per formula unit and c/a ratio for these compositions. The c/a ratio decreases sharply with increasing Zr-content and becomes very close to one for x≥0.545. This indicates that the nature of the monoclinic phase changes from pseudo-tetragonal to pseudo-rhombohedral around x ≈ 0.545. The unit cell volume increases sharply for x<0.545 and then increases slowly for higher Zr-content. In contrast Noheda et al [4] have reported that the unit cell volume changes linearly. Since Noheda et al [4] have used only a few reflections to determine the cell parameters, they have missed the abrupt change of the cell volume close to the MPB (i.e., x ≈ 0.520).

It is clear from the foregoing that the structure of PZT is pure monoclinic for composition range 0.525≤x≤0.60 and there is no rhombohedral phase as such on the Zr-rich side of the MPB. This supports the original idea of Ragini et al [16] which was further refined by Glazer et al [23] that the structure of PZT is short range monoclinic on the tetragonal side of the MPB (as $Pb^{2+}$ ions are displaced from cube corners along monoclinic <110> directions [6, 16]), becomes long ranged monoclinic in the MPB region as shown in the present work for x=0.525, and again becomes short ranged monoclinic for the Zr-rich compositions. The peak broadening due to short range ordered $M_A$ domains masks the observation of characteristic splitting of the peaks for x>0.525. In an independent work, Schönau et al [5] have also refined the structure of PZT using monoclinic Cm space group for x>0.575 in samples whose MPB is not only shifted to higher Zr-content but consists of a wider coexistence region (Δx ≈ 0.05) which is due to extrinsic reasons like off-stoichiometry and poor chemical homogeneity of the samples prepared by solid state route. Their TEM studies reveal microdomains of the parent tetragonal compositions alongwith miniaturized nanodomains. They have proposed that the structure of these nanodomains may be rhombohedral (for which surprisingly no evidence has been advanced) whose thin platelike morphology might be responsible for the anisotropic peak broadening of the 200 type reflections. However, as per the PZT phase diagram [14] the room temperature phase for x=0.60 originates directly from the cubic phase and so there is no possibility of miniaturization of the tetragonal microdomains for this composition. However, as already shown in Fig.3, even this composition shows huge anisotropic peak broadening of reflections like 200 which can be accounted for using Cm space group only and not by the R3m space group. Thus the origin of the anomalous peak broadening for x≥0.530 is not due to the presence of the miniaturized tetragonal domains but rather due to the presence of a lower symmetry monoclinic $M_A$ phase.

To summarize, using Rietveld analysis of powder synchrotron XRD data on stoichiometric and chemically homogeneous PZT samples, we have shown that a monoclinic phase in the Cm space group exists for 0.525≤ x ≤0.60 and that there is no evidence for a higher symmetry rhombohedral phase for such compositions.

**Figure Caption:**

**Fig.1** Powder synchrotron XRD profiles for 200, 220 and 222 pseudocubic reflections of PZT in the composition range 0.515 to 0.60.

**Fig.2** Observed (dots), calculated (continuous line) and difference (bottom line) profiles of 110, 111 and 200 pseudocubic reflections for $Pb(Zr_{0.525}Ti_{0.475})O_3$ using various structural models. The vertical tick marks above the difference profiles give the positions of the Bragg reflections.

**Fig.3** Variation of the ratio of the FWHM of the profiles of the 200 and 111 Bragg reflections for Zr-rich compositions of PZT.

**Fig.4** Observed (dots), calculated (continuous line) and difference (bottom line) profiles of 111, 200 and 220 pseudocubic reflections of PZT for x=0.535 (a), (b), (c), (d), and x=0.60 (e), (f) using various structural models. The vertical tick marks above the difference profiles give the positions of the Bragg reflections.

**Fig.5** Variation of the equivalent elementary perovskite (a) cell parameters and (b) cell volume and c/a ratio with composition.

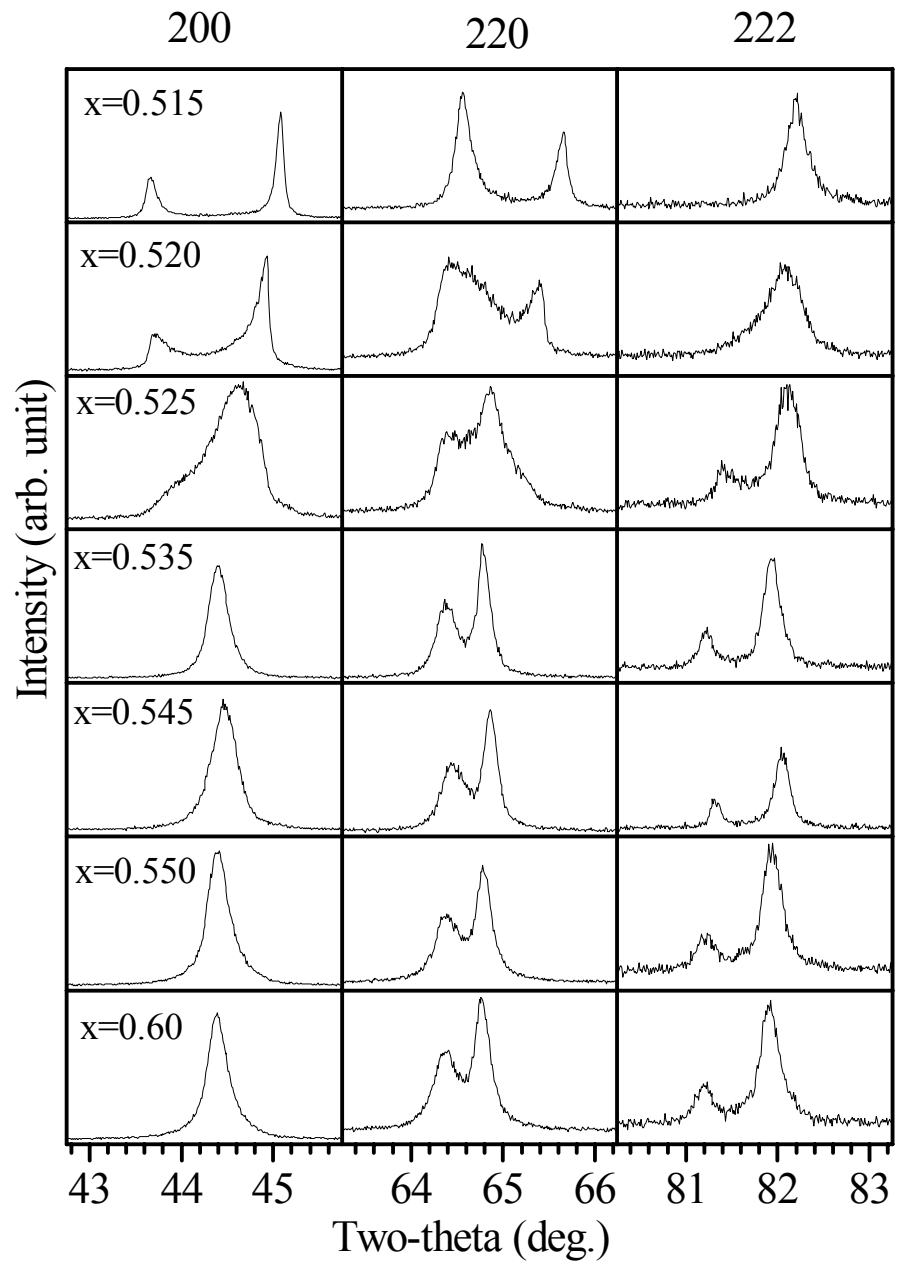

Fig.1

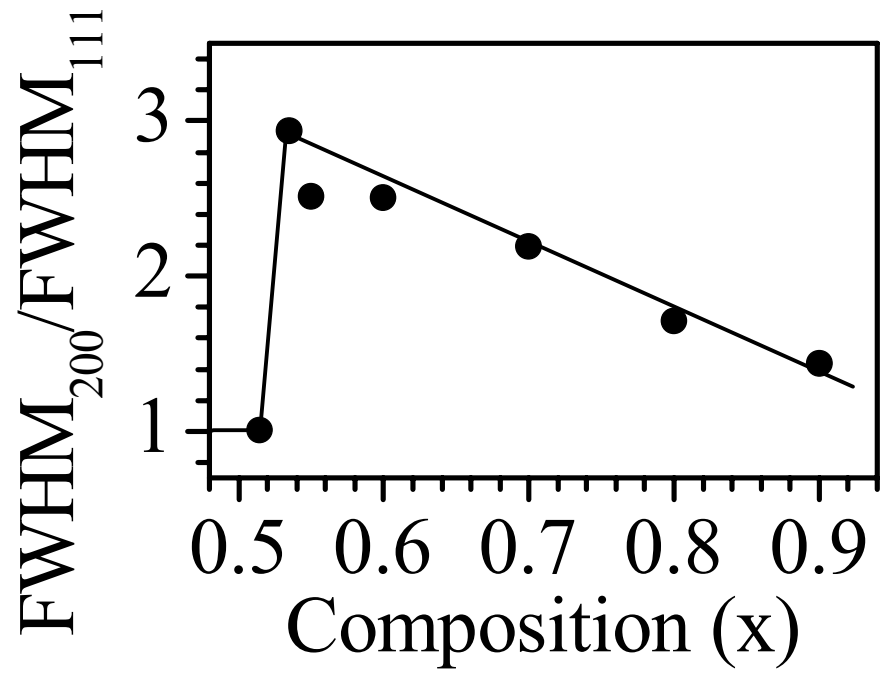

Fig.3

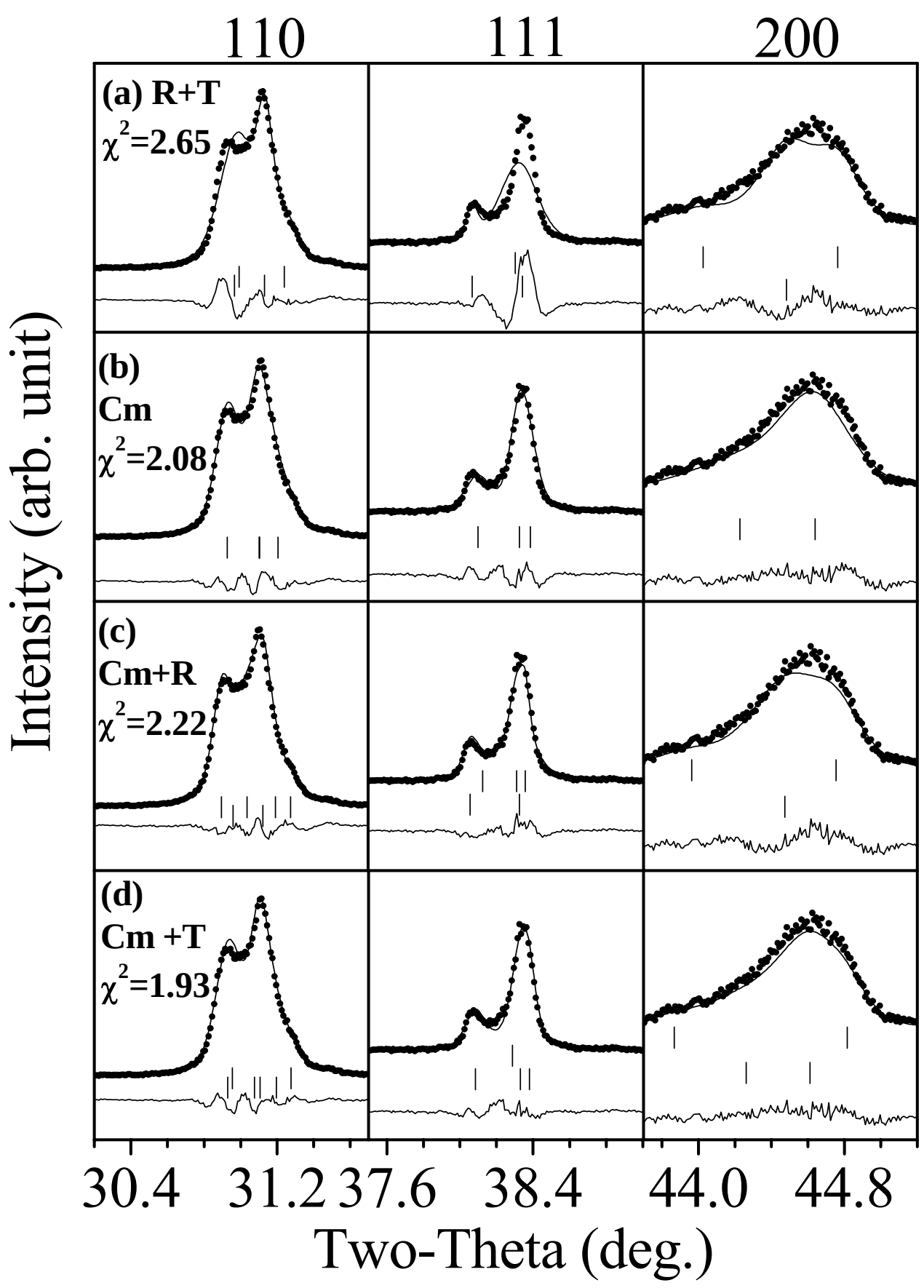

Fig.2

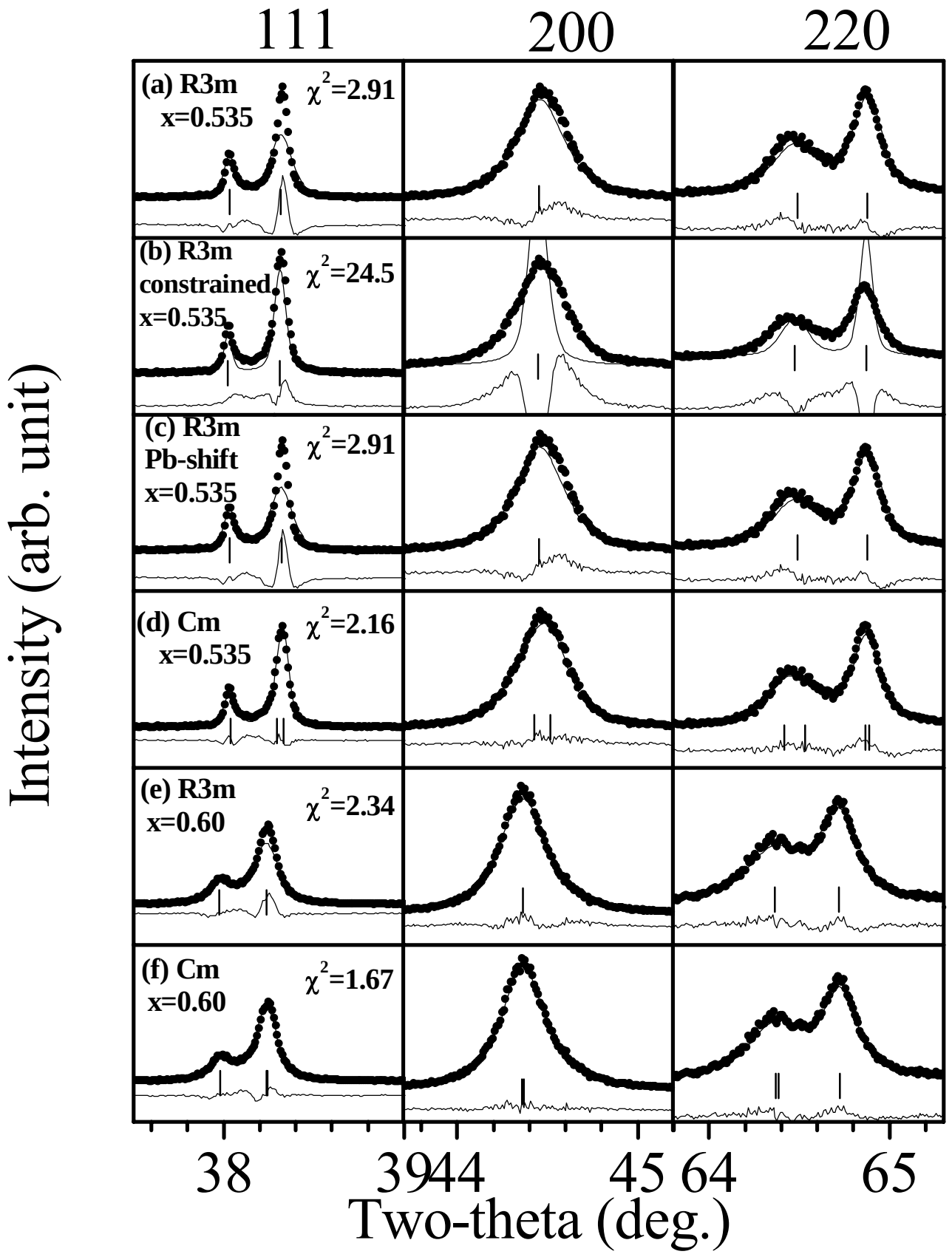

Fig.4

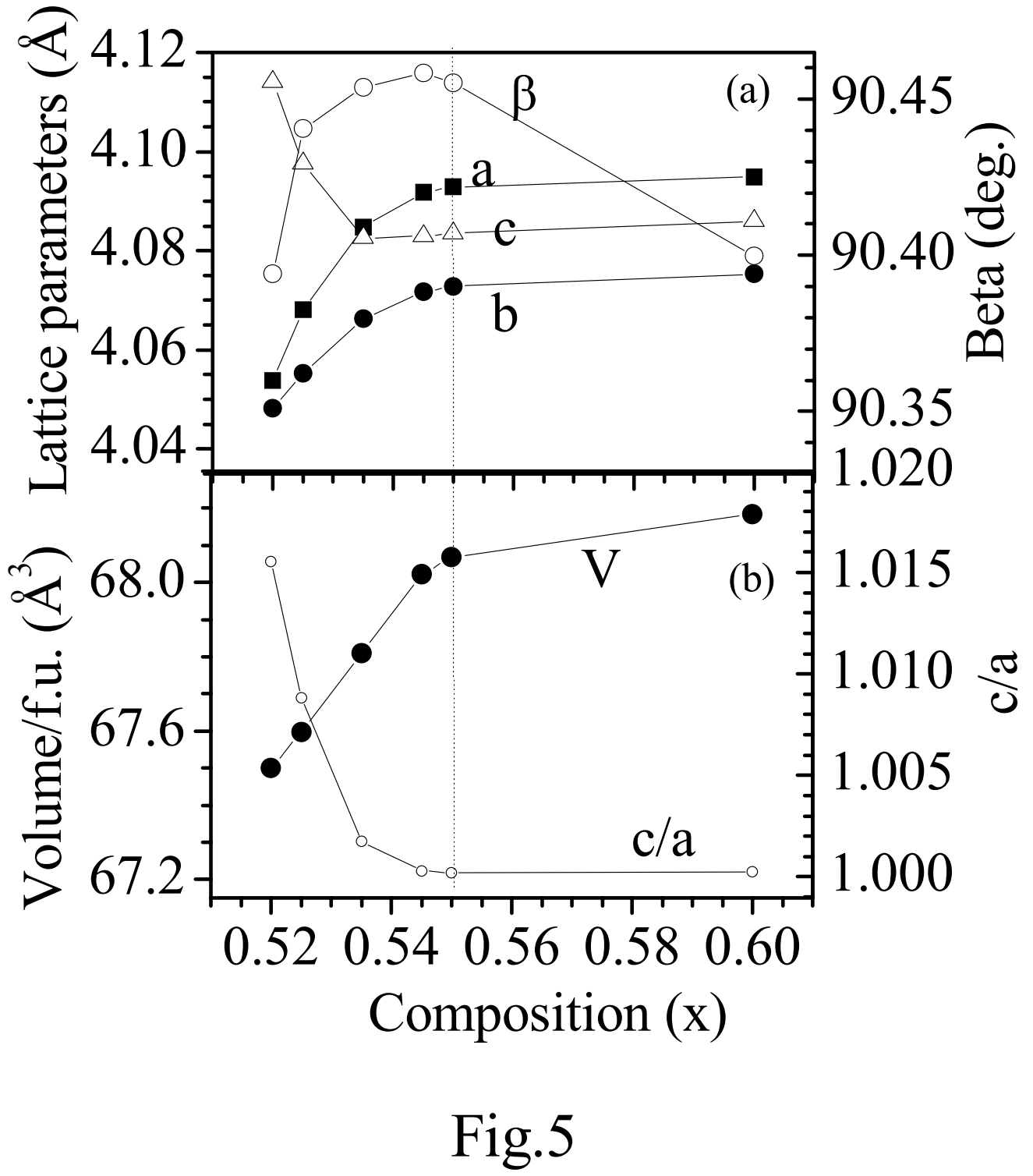

Fig.5